\documentclass[submission, Phys]{SciPost}
\usepackage[UKenglish]{babel}
\usepackage[utf8]{inputenc}
\usepackage[T1]{fontenc}

\usepackage{nicefrac}
\usepackage{amssymb}
\usepackage{cancel}

\newcommand{\tikzpic}[2]{
  \ifuseexttikz
  \includegraphics{#1}
  \else
  \tikzsetnextfilename{#1}
  #2
  \fi
}

\newif\ifuseexttikz
\useexttikztrue

\ifuseexttikz
\else
\usepackage{pgfplots}
\usepackage{tikz}
\usetikzlibrary{external} 
\usetikzlibrary{shapes}
\usetikzlibrary{calc}
\usetikzlibrary{external}
\fi

\hypersetup{pdfauthor={Claudius Hubig},pdftitle={Abelian and non-abelian symmetries in infinite projected entangled pair states}}

\begin{document}

\begin{center}{\Large \textbf{
Abelian and non-abelian symmetries in infinite projected entangled pair states
}}\end{center}

\begin{center}
Claudius Hubig
\end{center}

\begin{center}
Max-Planck-Institut für Quantenoptik,\\
Hans-Kopfermann-Str. 1, 85748 Garching, Germany
\\
claudius.hubig@mpq.mpg.de
\end{center}

\begin{center}
\today
\end{center}


\section*{Abstract} {\bf We explore in detail the implementation of
  arbitrary abelian and non-abelian symmetries in the setting of
  infinite projected entangled pair states on the two-dimensional
  square lattice. We observe a large computational speed-up; easily
  allowing bond dimensions $D=10$ in the square lattice Heisenberg
  model at computational effort comparable to calculations at $D=6$
  without symmetries. We also find that implementing an unbroken
  symmetry does not negatively affect the representative power of the
  state and leads to identical or improved ground-state
  energies. Finally, we point out how to use symmetry implementations
  to detect spontaneous symmetry breaking.}

\section{Introduction}
\label{sec:intro}

Tensor network methods have become the de facto method of choice in
the numerical treatment of one-dimensional quantum systems
\cite{schollwoeck11}. The extension to other loop-free geometries is
relatively straightforward\cite{shi06:_class, murg10:_simul,
  tagliacozzo09:_simul, gerster14:_uncon, gunst18:_t3ns,
  bauernfeind17:_fork_tensor_produc_states} and preserves most of the
numerical advantages of the one-dimensional systems. Work is still
ongoing, however, to effectively and efficiently handle large
two-dimensional systems which are firstly both of extremely high
experimental and theoretical interest and secondly difficult to treat
numerically with any method. For example, the dynamical mean-field
theory is being combined with new and different
solvers\cite{wolf15:_imagin_time_matrix_produc_state,
  stadler15:_dynam_mean_field_theor_plus,
  bauernfeind17:_fork_tensor_produc_states} to allow the treatment of
larger clusters within the dynamical cluster approximation; a similar
approach is taken for the density matrix embedding
theory\cite{knizia12:_densit_matrix_embed,knizia13:_densit_matrix_embed}
to likewise allow for larger clusters\cite{zheng17:_strip_hubbar}
potentially necessary for the treatment of two-dimensional
lattices. Recently, a diagrammatic formulation of quantum Monte
Carlo\cite{deng15:_emerg_bcs_hubbar} was also used to further avoid
the sign problem, which otherwise makes calculations in frustrated or
fermionic systems very hard. Standard matrix-product
state/density-matrix renormalisation group (MPS-DMRG) methods are also
continuously applied to wider and wider cylinders, but computational
costs essentially explode beyond relatively limited cylinder
circumferences\cite{stoudenmire12:_study_two_dimen_system_densit,
  leblanc15:_solut_two_dimen_hubbar_model, ehlers17:_hybrid_hubbar,
  depenbrock12:_natur_spin_liquid_groun_state}.

Projected entangled pair states \cite{verstraete04:_renor}, the
two-dimensional generalisation of matrix-product states, were proposed
soon after the understanding of the latter in the context of the
density matrix renormalisation group and a further generalisation to
\emph{infinite} two-dimensional systems\cite{verstraete08:_matrix,
  jordan08:_class_simul_infin_size_quant} followed with continuous
improvements\cite{corboz10:_simul, corboz10:_simulb,
  corboz14:_compet_states_model_b, lubasch14:_algor, phien15:_fast,
  phien15:_infin, vanderstraeten16:_gradien, corboz16:_variat,
  corboz16:_improv_hubbar} to the method over the last few years. In
this work, we explore the implementation of both abelian
and non-abelian symmetries -- which is a standard method in
one-dimensional MPS codes -- with the infinite projected entangled
pair states (iPEPS) method used to calculate two-dimensional
ground-state properties. In particular, we will show that, absent
spontaneous symmetry breaking, implementing an abelian or non-abelian
symmetry provides a major computational advantage and considerably
improved estimates for the ground-state energy. This will be done
using the explicit Clebsch-Gordan Coefficients approach introduced in
Ref.~\cite{weichselbaum12:_non}. Furthermore, we will showcase how the
implementation of a symmetry allows additional control over a
calculation: First, once the symmetry is implemented, one forces the
calculation into a canonical ensemble, hence exactly preserving
e.g.~particle number at the desired and manually-selected
value. Second, enabling or disabling the preservation of the symmetry
allows selecting between symmetry-unbroken and symmetry-broken states
and phases. By comparing the energies of these states, it becomes very
easy and very stable to pinpoint phase transitions.

To this end, the paper is structured as follows: Sec.~\ref{sec:syms}
reviews the implementation of arbitrary symmetries in tensor networks
and establishes some notation. Sec.~\ref{sec:impl-details} goes into
further detail of required adaptations of the standard iPEPS
algorithms to allow for such an implementation as well as different
considerations of computational effort. In Sec.~\ref{sec:comp-speed},
we use the Heisenberg $S=\nicefrac{1}{2}$ Hamiltonian on a square and
Kagomé lattice to show the computational speed-up attainable by
implementing symmetries in the network. In particular, a large benefit
due to the implementation of the unbroken $\mathrm{SU}(2)$-spin
symmetry in the Kagomé lattice is observed. The selection of
symmetry-broken and symmetry-unbroken states and their use in
diagnosing spontaneous symmetry breaking is discussed in
Sec.~\ref{sec:sym-breaking}. Finally, we conclude in
Sec.~\ref{sec:conclusions}.

\section{\label{sec:syms}Symmetries in tensor networks}

The implementation of global abelian symmetries in low-rank loop-free
tensor networks is generally well-understood and a cornerstone of
efficient computations in modern codes\cite{schollwoeck11,
  hauschild:_tensor_networ_python_tenpy, mcculloch16:_mp_toolk,
  stoudenmire:_itens_c}. In contrast to this, global non-abelian
symmetries\cite{mcculloch02:_abelian} have seen less widespread use
and proposals on how to implement them for higher-rank tensor networks
are relatively new \cite{weichselbaum12:_non, singh12:_tensor_su}.
While global abelian symmetries generalise straightforwardly to
higher-rank tensors (see e.g.~Ref.~\cite{bauer11:_implem_abelian,
  haghshenas18:_j_heisen} for applications to (i)PEPS), for
non-abelian symmetries there are two alternative approaches: the
\emph{implicit Clebsch-Gordan coefficients approach} relies on
analytical knowledge of Clebsch-Gordan coefficients (CGCs) for rank-3
tensors and decomposes any higher-rank tensor into a double tree of
rank-3 tensors \cite{singh10:_tensor,
  singh12:_tensor_su}. Alternatively, the \emph{explicit CGC approach}
generalises the notion of CGCs to higher-rank tensors and explicitly
stores those coefficients within the tensor
\cite{weichselbaum12:_non}. The latter approach was already applied to
iPEPS calculations in Ref.~\cite{bruognolo17:_tensor, liu15:_simpl_heisen}.  

Here, we will briefly summarise the implementation of arbitrary
symmetries following the \emph{explicit CGC approach} as done in the
\textsc{SyTen} toolkit \cite{hubig:_syten_toolk}. For a more extensive
discussion of this approach, see Refs.~\cite{weichselbaum12:_non,
  hubig17:_symmet_protec_tensor_networ}. In
Sec.~\ref{sec:impl-details} we will then go on to discuss
implementation details specific to iPEPS calculations.

Fundamentally, a rank-$R$ tensor $T$ is represented as a direct sum of
a finite number of blocks $\{ t_i \}$:
\begin{equation}
  T = \bigoplus_{i=1}^{N_t} t_i \quad. \label{eq:tensor-blocks}
\end{equation}
Each of the $R$ legs of the tensor is given a direction corresponding
to the tensor acting on the vector space on this leg or its dual. A
matrix, for example, would have one incoming leg and one outgoing
leg. Every block $t_i$ then firstly contains a single quantum number
label for every symmetry in the system (e.g.~$\mathrm{U}(1)$ particle
number $n$ and $\mathrm{SU}(2)$ spin $S$) on each of its $R$
legs. Furthermore, each block contains one dense \emph{reduced}
rank-$R$ tensor $m_i$ and, for every symmetry $\gamma$, a
\emph{generalised CGC} rank-$R$ tensor $c^\gamma_i$:
\begin{equation}
  t_i = m_i \bigotimes_{\gamma=1}^{N_S} c^\gamma_i \quad. \label{eq:tensor-cgc}
\end{equation}
Blocks with the same quantum numbers on all legs can be added
directly, blocks with different quantum numbers are placed next to
each other (cf.~Eq.~\eqref{eq:tensor-blocks-diag}). If there are no
symmetries in the system, the tensor $T$ only contains a single block
$t_1$ which, in turn, only contains the collection of scalar
coefficients of the tensor. For abelian symmetries, $c^\gamma_i$ are
one-dimensional tensors which are non-zero if the addition rules for
the abelian symmetry are fulfilled by the quantum numbers assigned to
the tensor legs and zero if the addition rules are
broken. Alternatively, the $c^\gamma_i$ can be left off and the
addition rules enforced by hand. For rank-2 tensors with one incoming
and one outgoing index, $c^\gamma_i$ are proportional to identity
matrices, for rank-3 tensors, they correspond directly to the standard
Clebsch-Gordan coefficients.

\paragraph{As an example,} consider a rank-2 tensor $T$ with a single
preserved $\mathrm{SU}(2)$-spin symmetry. The blocks of the tensor are
labelled by the $\mathrm{SU}(2)$-spin quantum numbers of the states on
the incoming and outgoing tensor legs. Let the tensor contain three
blocks corresponding to $S=0$, $S=\nicefrac{1}{2}$ and $S=1$:
\begin{equation}
  T = t_{0,0} \bigoplus t_{\nicefrac{1}{2},\nicefrac{1}{2}} \bigoplus t_{1,1} \quad.
\end{equation}
The implicit direct sum is equivalent to placing the three blocks on
the diagonal of a larger matrix:
\begin{equation}
  T = \begin{pmatrix} \left[ t_{0,0} \right] & 0 & 0 \\ \xcancel{t_{\nicefrac{1}{2},0}} & \left[ t_{\nicefrac{1}{2},\nicefrac{1}{2}} \right] & 0 \\ 0 & 0 & \left[ t_{1,1} \right] \end{pmatrix} \quad. \label{eq:tensor-blocks-diag}
\end{equation}
The location of a hypothetical block $t_{\nicefrac{1}{2},0}$ is also
shown but this block (like all other zeros in the $T$ tensor) is
forbidden by the symmetry and must be identically zero. The
decomposition according to Eq.~\eqref{eq:tensor-cgc} then allows
rewriting each block $t_i$ as the tensor product of a reduced block
and a symmetry-protected CGC tensor $c_i$. Since the tensor $T$ is a
rank-2 tensor, the $c_i$ are simply identity matrices of the
appropriate size: $1 \times 1$, $2 \times 2$ and $3 \times 3$
respectively, as the irreducible representations $S=0$,
$S=\nicefrac{1}{2}$ and $S=1$ are one-, two- and three-dimensional
respectively:
\begin{align}
  t_{0,0} & = m_{0,0} \bigotimes \mathbf{1}_1 \\
  t_{\nicefrac{1}{2},\nicefrac{1}{2}} & = m_{\nicefrac{1}{2},\nicefrac{1}{2}} \bigotimes \mathbf{1}_2 \\
  t_{1,1} & = m_{1,1} \bigotimes \mathbf{1}_3 \quad.
\end{align}
While $t_{0,0}$ and $m_{0,0}$ have exactly the same number of rows and
columns, $t_{\nicefrac{1}{2},\nicefrac{1}{2}}$ has twice the number of
rows and columns as $m_{\nicefrac{1}{2},\nicefrac{1}{2}}$ (and four
times the number of elements) while $t_{1,1}$ has nine times as many
elements as $m_{1,1}$! This already is an indicator of the power of
non-abelian symmetries in tensor networks.

In comparison, an \emph{abelian} symmetry would still allow for the
decomposition of the tensor $T$ into blocks. The $S^z = 0$ block
$t_{0,0}$, however, would contain contributions from the $S=0$ and the
$S=1$ sector (and hence would be larger). It would also not be
possible to exploit the two-fold degeneracy of the $S=\nicefrac{1}{2}$
sector, one would instead have to keep two blocks
$t_{\pm \nicefrac{1}{2}}$. In this case, all $c_i$ would be
$1 \times 1$ dummy matrices.

Note that here we force every tensor to preserve the symmetry exactly
and do not allow for a shift of all quantum numbers within a tensor,
as such shifts are only well-defined for abelian symmetries and would
break down immediately when including non-abelian symmetries.

\paragraph{Tensor operations} can then be implemented to largely act
independently on each block and also the $m_i$ and $c^\gamma_i$ within
each block. The addition of two tensors $T + S$ is simply a direct sum
over all blocks:
\begin{equation}
  T + S = \bigoplus_{x=t,s} \bigoplus_{i=1}^{N_x} x_i \quad.
\end{equation}
However, we may then end up with many blocks with identical quantum
numbers on all legs and an identical Clebsch-Gordan structure. If for
every symmetry in the system, the associated CGC tensors of two blocks
are parallel to each other, we may move those factors into the reduced
dense tensors and then add the reduced dense tensors together to
arrive at a single block consisting of the sum of the dense tensors as
well as one set of CGC tensors. This operation hence reduces the
number of blocks in the direct sum and is called a \emph{reduction} in
the following (not to be confused with the reduced tensor $m_i$
above).

The contraction of two tensors $T$ and $S$, $R = T \cdot S$ over a set
of legs is evaluated as follows: First, collect all pairs of blocks
$\{ t_i, s_j \}$ which are labelled by identical quantum numbers on
the contracted legs. Then construct result blocks $r_k$ as
\begin{equation}
  r_k = \left( m_{t,i} \cdot m_{s,j} \right) \bigotimes_{\gamma=1}^{N_s} \left( c^\gamma_{t,i} \cdot c^\gamma_{s,j} \right)
\end{equation}
where $\cdot$ denotes the desired contraction. Afterwards reduce the
number of blocks of the result tensor $R$ as described above by adding
parallel tensor blocks.

While it is straightforward to parallelise over the pairs of all
possible contractions, to reduce memory usage, one should also first
collect all contraction pairs resulting in a given quantum number
configuration on the open legs, contract and reduce those and only
then move on to the next set of quantum numbers on the open
legs. Otherwise, the intermediate, unreduced $R$ tensor can easily
exceed the size of the actual reduced $R$ by a factor of 100.

\paragraph{Reduced and Total Dimensions} When speaking about the
\emph{bond dimension} of a tensor network state or more generally the
dimensions of a tensor, it is useful to differentiate between the
\emph{reduced} and the \emph{total} dimension. Here, the \emph{reduced
  dimension} of a tensor on one of its legs is the sum of the sizes of
every reduced block (counted only once per quantum number sector). In
comparison, the \emph{total dimension} of a tensor on one of its legs
is the sum of the sizes of every reduced block multiplied by the
product of the sizes of the associated generalised CGC tensors. Put
another way, a spin doublet state with two duplicate singular values
associated to $\pm \nicefrac{1}{2}$ has a total dimension of two
regardless of the symmetries used to represent it. However, if the
tensor implements the $\mathrm{SU}(2)$ spin symmetry, the
\emph{reduced} dimension of the state is merely $1$, the
duplicate-ness of the singular values is not counted. We then
typically expect computational effort to scale cubically in the
reduced dimensions, whereas the expressivity of a tensor and the
potentially carried entanglement scales with its total dimensions. To
differentiate the two in the following text, we will use capital
letters ($D$, $X$) to refer to total dimensions whereas lower-case
letters ($d$, $\chi$) refer to the reduced dimensions. In the iPEPS
calculations to follow, $d$ and $D$ respectively will denote the state
bond dimension whereas $\chi$ and $X$ respectively will denote the
environment bond dimension.

The distribution of states among the different quantum number sectors
is in principle open and should be adapted according to the specific
state represented. While at relatively small bond dimensions (as
encountered in iPEPS calculations) it may be feasible to fix that
distribution by hand beforehand, finding the optimal choice a priori
becomes combinatorially expensive. To avoid this problem, minimal care
must be taken to allow flexibility in the choice of quantum number
sectors. Within the full update scheme with imaginary time evolution,
this flexibility comes naturally. For the newer variational update
schemes \cite{vanderstraeten16:_gradien, corboz16:_variat} one may
have to consider either two-site updates or an expansion scheme
analogous to the subspace expansion in DMRG
\cite{hubig15:_stric_dmrg}.

\section{Implementation details specific to iPEPS}
\label{sec:impl-details}

Our iPEPS calculation largely follows the existing literature, making
use of imaginary time evolution with the fast full update
(FFU)\cite{phien15:_fast} and gauge fixing\cite{lubasch14:_algor,
  phien15:_infin}. The corner transfer matrix is truncated as
suggested in \cite{corboz14:_compet_states_model_b}. During the
optimisation, we aim for a total environment bond dimension $X$
approximately ten times the total state bond dimension $D$.

Errors for observables from the corner transfer matrix method are
estimated as follows: First, at given environment bond dimension
$\chi$, corner transfer matrix (CTM) steps are done until the
observable changes by less than a small threshold ($10^{-8}$) or 15
steps are done. The first error is then the maximum change between the
last three steps, the last value is taken as the result of this
fixed-$\chi$ calculation. Second, $\chi$ is multiplied by 2 and the
previous procedure repeated until the resulting values differ by less
than the small threshold $10^{-8}$ or $\chi$ is larger than 128. The
second error is then the change between the value at the last fixed
$\chi$ and the second-to-last calculation at fixed $\chi$. The error
is the larger of the first and second error. In our experience, this
is a very conservative error measure which is likely to overestimate
the error of the obtained value. Absent a better error estimation
method, we will however use it in the following figures. 

While the variational update is not yet implemented, its improvements
are orthogonal to those discussed here, so we expect all results to
carry over from imaginary time evolution to the variational update
with minimal modifications. The same applies to the new extrapolation
schemes for observables\cite{corboz16:_improv_hubbar,
  rader18:_finit_correl_lengt_scalin_loren}.

Additionally, we should differentiate this tensor-based approach to
global symmetries from an alternative in which one first parametrises
all (e.g.)  $\mathrm{SU}(2)$ invariant tensors with a specific bond
dimension and specific quantum numbers on each bond. In this
alternative approach, one then works with dense state tensors (with
complicated parametrised entries, which are optimised) but the
symmetries do not necessarily extend to (e.g.) the environment
tensors. This approach is in particular useful if the obtainable bond
dimension is relatively small, as one may categorise and understand
each individual state at a specific bond dimension and the relative
advantage of a symmetry implementation at very small bond dimensions
is also small.

\subsection{Required adaptations of the iPEPS setup}

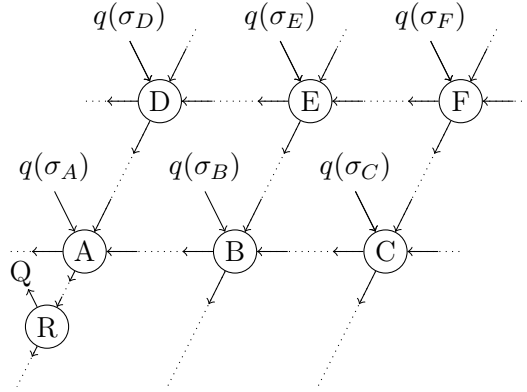
\begin{figure}[tb]
  \centering
  \tikzpic{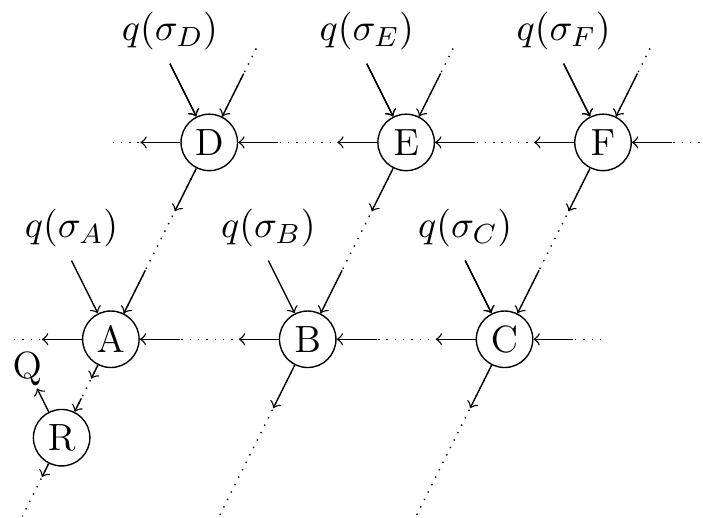}{
    \begin{tikzpicture}

      \node (A) [draw,circle,inner sep=0pt, minimum size=1.5em] at (0,0) {A};
      \node (B) [draw,circle,inner sep=0pt, minimum size=1.5em] at (2,0) {B};
      \node (C) [draw,circle,inner sep=0pt, minimum size=1.5em] at (4,0) {C};
      \node (D) [draw,circle,inner sep=0pt, minimum size=1.5em] at (1,2) {D};
      \node (E) [draw,circle,inner sep=0pt, minimum size=1.5em] at (3,2) {E};
      \node (F) [draw,circle,inner sep=0pt, minimum size=1.5em] at (5,2) {F};
      \node (R) [draw,circle,inner sep=0pt, minimum size=1.5em] at (-0.5,-1) {R};

      \draw[<-] (A) -- +(0.7,0);
      \draw[->] (A) -- +(-0.7,0);
      \draw[<-] (A) -- +(0.35,0.7);
      \draw[->] (A) -- +(-0.2,-0.4);
      \draw[<-] (A) -- +(-0.4,0.8) node[above] {$q(\sigma_A)$};
      
      \draw[<-] (B) -- +(0.7,0);
      \draw[->] (B) -- +(-0.7,0);
      \draw[<-] (B) -- +(0.35,0.7);
      \draw[->] (B) -- +(-0.35,-0.7);
      \draw[<-] (B) -- +(-0.4,0.8) node[above] {$q(\sigma_B)$};

      \draw[<-] (C) -- +(0.7,0);
      \draw[->] (C) -- +(-0.7,0);
      \draw[<-] (C) -- +(0.35,0.7);
      \draw[->] (C) -- +(-0.35,-0.7);
      \draw[<-] (C) -- +(-0.4,0.8);
      \draw[<-] (C) -- +(-0.4,0.8) node[above] {$q(\sigma_C)$};

      \draw[<-] (D) -- +(0.7,0);
      \draw[->] (D) -- +(-0.7,0);
      \draw[<-] (D) -- +(0.35,0.7);
      \draw[->] (D) -- +(-0.35,-0.7);
      \draw[<-] (D) -- +(-0.4,0.8);
      \draw[<-] (D) -- +(-0.4,0.8) node[above] {$q(\sigma_D)$};

      \draw[<-] (E) -- +(0.7,0);
      \draw[->] (E) -- +(-0.7,0);
      \draw[<-] (E) -- +(0.35,0.7);
      \draw[->] (E) -- +(-0.35,-0.7);
      \draw[<-] (E) -- +(-0.4,0.8);
      \draw[<-] (E) -- +(-0.4,0.8) node[above] {$q(\sigma_E)$};

      \draw[<-] (F) -- +(0.7,0);
      \draw[->] (F) -- +(-0.7,0);
      \draw[<-] (F) -- +(0.35,0.7);
      \draw[->] (F) -- +(-0.35,-0.7);
      \draw[<-] (F) -- +(-0.4,0.8);
      \draw[<-] (F) -- +(-0.4,0.8) node[above] {$q(\sigma_F)$};

      \draw[->] (R) -- +(-0.2,-0.4);
      \draw[<-] (R) -- +(0.2,0.4);
      \draw[->] (R) -- +(-0.25,0.5) node at +(-0.35,0.7) {Q};

      \draw [dotted] (A) -- (D);
      \draw [dotted] (A) -- (B);
      \draw [dotted] (B) -- (C);
      \draw [dotted] (B) -- (E);
      \draw [dotted] (C) -- (F);
      \draw [dotted] (D) -- (E);
      \draw [dotted] (E) -- (F);

      \draw [dotted] (A) -- (R);
      
      \draw [dotted] (A) -- +(-1,0);
      \draw [dotted] (D) -- +(-1,0);
      \draw [dotted] (C) -- +(+1,0);
      \draw [dotted] (F) -- +(+1,0);

      \draw [dotted] (D) -- +(0.5,1);
      \draw [dotted] (E) -- +(0.5,1);
      \draw [dotted] (F) -- +(0.5,1);
      
      \draw [dotted] (B) -- +(-0.9,-1.8);
      \draw [dotted] (C) -- +(-0.9,-1.8);
      \draw [dotted] (R) -- +(-0.4,-0.8);
    \end{tikzpicture}
  }
  \caption{\label{fig:remover}Layout of an example $2 \times 3$ iPEPS
    unit cell as used here. Each of the individual site tensors $A$,
    $B$, $C$, $D$, $E$ and $F$ inserts some charge into the network
    (which also depends on the local state), the special remover
    tensor $R$ is used to extract a total charge $Q$ from the unit
    cell. Auxiliary leg contractions are indicated by the dotted
    lines. The average quantum number on the inter-unit-cell auxiliary
    legs is zero (with some fluctuations) whereas intra-unit-cell legs
    can have non-zero average quantum number depending on the values
    of $q(\sigma)$ and the ``routing'' of charges through the unit
    cell.}
\end{figure}

\label{sec:impl-details:adaptations}
The first step to combine symmetry-preserving tensors with an
established tensor network algorithm is to define a consistent leg
direction such that the tensor contractions required to, for example,
represent the state are well-defined. The specific leg directions are
entirely arbitrary, but of course have to be consistent within a
calculation.\footnote{Even when not enforcing the symmetry at the
  tensor level, a consistent choice of directions is required, as
  observed e.g.~in Ref.~\cite{hackenbroich18:_inter_peps}.}

Second, when enforcing the symmetry, we effectively work in a
canonical ensemble of a fixed average quantum number per unit cell. If
this average quantum number is non-zero (e.g. 1 particle per site in
the Hubbard model), it shifts the outgoing auxiliary tensor legs
relative to the incoming auxiliary tensor legs. To account for this
shift and give an overall consistent state, we insert an additional
rank-3 \emph{remover} tensor on the boundary of each unit cell. The
remover tensor has a ``physical'' leg which points in the opposite
direction of the usual physical legs on our iPEPS sites and simply
removes as many particles as desired from the calculation. The
additional leg is traced out during all calculations, for abelian
symmetries, it amounts to an additional shift of the quantum numbers
in each unit cell. The remover tensor is initialised as a random
tensor with the required quantum number sectors on its three legs and
then optimised during the iPEPS update of its two neighbouring sites.

Third, when constructing a single-site operator such as $\hat s^+$ in
a $\mathrm{U}(1)$-$S^z$ preserving system, it is necessary to insert
an additional tensor leg to carry the additional quantum number. In
the product $\hat s^+_i \hat s^-_j$, these legs then combine to form a
two-site gate acting on sites $i$ and $j$ connected by the contracted
leg. If the contracted leg crosses any other tensor legs, the usual
rule for fermionic PEPS applies\cite{corboz10:_simul} and a swap gate
has to be inserted\cite{corboz10:_simulb}. This swap gate fulfills the
same role as Jordan-Wigner strings between operators $\hat c^\dagger$
and $\hat c$ in matrix-product operators \cite{hubig17:_gener}.

\subsection{Performance and accuracy necessitated by iPEPS compared to MPS}

In a typical MPS-DMRG calculation, the maximal ranks of temporary
tensors is typically limited to four or five, with the basic state
building blocks having rank three. This has two beneficial side
effects: first, the number of blocks in a tensor is limited to
approximately $O(1000)$ with each individual block being relatively
large. Second, the entries of the CGC tensors are regularly
constructed from rank-2 and rank-3 tensors, which helps alleviate the
buildup of numerical errors due to limited accuracy.

In an iPEPS calculation, in contrast, the maximal tensor ranks are
much larger (rank 10 or above!), naively leading to tensors with
$O(10^6)$ or more blocks. Furthermore, numerical errors in the CGC
tensors build up very quickly as the basic rank-5 tensors already
display a complicated Clebsch-Gordan structure.

The first issue is solely a performance problem and can be avoided by
minimising tensor ranks wherever possible. In particular, it turns out
that first constructing the rank-4 double-layer tensors and then
applying the CTM algorithm based on the rank-4 tensors is effectively
computationally more efficient than contracting the single-layer
rank-5 tensors individually into the existing CTM tensors, even though
the latter approach is asymptotically cheaper. However, the former
decreases the overhead both from bookkeeping the tensors and
initialising many extremely small matrix-matrix products. As iPEPS
bond dimensions $D$ are typically relatively small, argument about
asymptotic efficiency also carry less weight than in (e.g.) MPS
methods.

The second problem is more intricate and delayed by three tricks:
First, we use higher-precision floating point
numbers\footnote{Specifically, the
  \texttt{boost::multiprecision::cpp\_bin\_float<76>} type is used.} to
represent the entries of the CGC tensors. Second, within a tensor, we
regularly check all non-zero coefficients (of which there are
typically very few) and set those coefficients which are close in
magnitude equal to each other. As an illustrative example, if a CGC
tensor contains coefficients $[0.99, 1.01, -1.01, 0.99]$, we set them
to $[1.00, 1.00, -1.00, 1.00]$. The actual closeness threshold used is
$\approx 10^{-16}$. This relies on the fact that the coefficients of a
CGC tensor are computed from very few inputs in very specific ways and
no (typical) CGC tensor should truly contain coefficients different
from each other but closer than $10^{-16}$. Finally, it is helpful to
regularly reconstruct the CGC tensors from scratch to remove all
numerically buildup errors. This is done by first merging all tensor
legs of the tensor until one arrives at a tensor with a single
incoming and a single outgoing leg. In this tensor, the CGC tensors
should be proportional to identity matrices which is then ensured
manually. Afterwards, the tensor legs are split again to rebuild the
original tensor structure.

An alternative solution for the issue of numerical accuracy would be
to use continued fractions to represent the CGC coefficients to high
accuracy as suggested in Ref.~\cite{weichselbaum12:_non,
  bruognolo17:_tensor} at the cost of a higher effort to
implementation.

With these relatively minor changes, it becomes possible to do
$\mathrm{SU}(2)$- and $\mathrm{U}(1)$-symmetric calculations on
infinite projected entangled pair states using symmetric tensors in
place of standard dense tensors without internal structure.

\section{Computational speed-up}\label{sec:comp-speed}

In classical DMRG calculations, implementing an abelian
$\mathrm{U}(1)$ symmetry typically leads to an approximately ten-fold
computational speed-up or, alternatively, an approximately tripled
bond dimension. Changing from a $\mathrm{U}(1)$ to a $\mathrm{SU}(2)$
symmetry has a similar effect of approximately tripling the effective
bond dimension. To test the effect of the implemented symmetries, we
pick two example systems, namely $S = \nicefrac{1}{2}$ spins on the
Kagomé and square lattice.

\subsection{Spin-\nicefrac{1}{2} square lattice Heisenberg model}

As a first example, we consider the $S=\nicefrac{1}{2}$ Heisenberg
Hamiltonian on a square lattice at the Heisenberg point of isotropic
coupling:
\begin{equation}
  \hat H = \sum_{\langle i, j \rangle} \hat s^x_i \hat s^x_j + \hat s^y_i \hat s^y_j + \hat s^z_i \hat s^z_j \quad.
\end{equation}
The system spontaneously breaks the $\mathrm{SU}(2)$ symmetry of the
Hamiltonian but preserves both the $\mathbb{Z}_2$ and
$\mathrm{U}(1)$-$S^z$ symmetries.

\paragraph{The computational speed-up} obtainable is first estimated
by measuring the time required to re-initialise the corner transfer
matrix with eight steps in each direction, evaluate the energy
(evaluating eight two-site expectation values) and then do five full
updates for imaginary time evolution of a second-order Trotter
decomposition on a 2x2 unit cell. Each step requires the application
of 16 gates. 8 of the 16 gates connect two neighbouring unit cells and
hence also require a directional (up/down or left/right) CTM step. The
runtime measurement is done after sufficiently many initial steps that
the iPEPS bond dimension $d$ and the bond dimension $\chi$ of the
environment has saturated. For abelian (or no) symmetries, we set
$\chi = 10(d+1)$, while for the non-abelian $\mathrm{SU}(2)$ case,
$\chi = 5(d+1)$ appears to be enough to obtain stability and
convergence.

\begin{figure}[tb]
  \begin{minipage}[t]{0.49\textwidth}
    \includegraphics{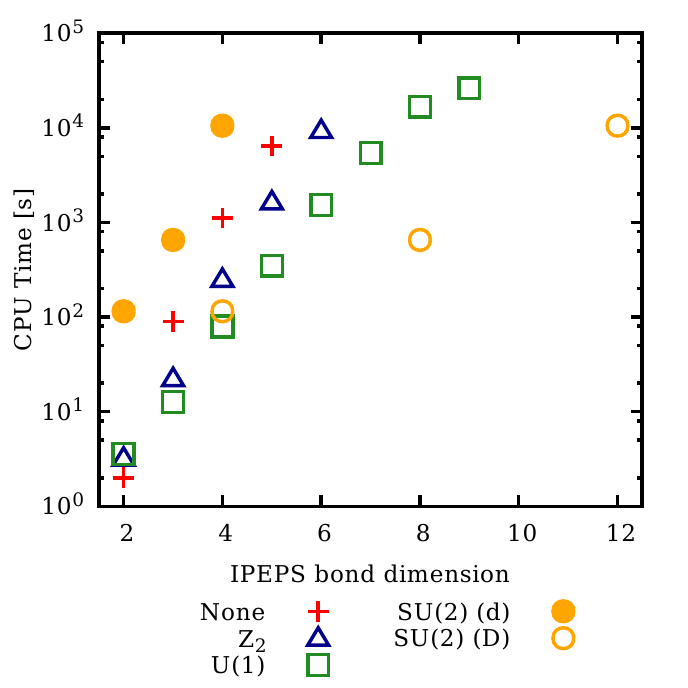}
    \caption{\label{fig:sq:effort-d}Computational effort for CTM
      initialisation, energy evaluation and five FFU steps in
      imaginary time in the Heisenberg square lattice model. Data for
      the $\mathrm{SU}(2)$-symmetric calculation is plotted over the
      reduced ($d$) and total ($D$) bond dimension.}
  \end{minipage}
  \hspace{0.02\textwidth}
  \begin{minipage}[t]{0.49\textwidth}
  \includegraphics{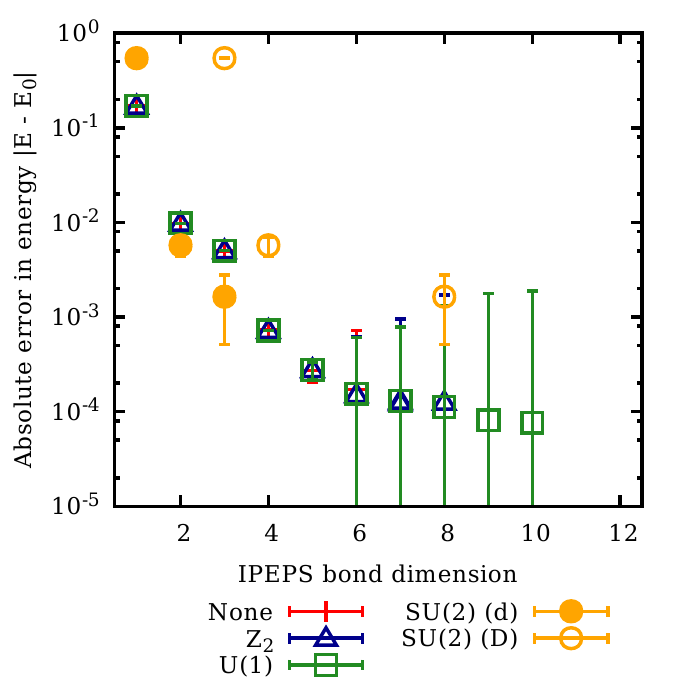}
  \caption{\label{fig:sq:error-d}Error in energy per spin calculated
    in the Heisenberg square lattice model at $\Delta = 1$ with
    $E_0 = -0.669437$. Error bars are obtained from the CTM
    calculation and give the error of the observable relative to its
    true value in the given variational state.}
  \end{minipage}
\end{figure}

Fig.~\ref{fig:sq:effort-d} shows the CPU time in seconds for the above
procedure. These measurements were done on a single core of a Intel
Xeon CPU E5-2630 v4 clocked at 2.20GHz. Except at minimal bond
dimension $D=2$, enforcing the $\mathrm{U}(1)$ or $\mathbb{Z}_2$
symmetry has a very beneficial effect on the runtime. At larger bond
dimensions, the $\mathrm{U}(1)$-symmetric calculation is also much
faster than the $\mathbb{Z}_2$-symmetric calculation. Comparing the
reduced bond dimension $d$ of a $\mathrm{SU}(2)$-symmetric
calculation, one observes a very large cost (e.g. at $d=3$: 89, 22, 12
and 653 seconds respectively without symmetries, with $\mathbb{Z}_2$,
$\mathrm{U}(1)$ and $\mathrm{SU}(2)$). Part of this large cost is due
to a relatively large overhead of the CGC tensors, but another part
comes from intermediate tensors (e.g. the rank-4 double layer tensors)
having a larger relative bond dimension which is less amenable to a
reduction in size by the symmetry implementation. On the other hand,
when comparing the total effective bond dimension $D$, the
$\mathrm{SU}(2)$-symmetric implementation is always faster (at
$D=4, 8, 12$) than a comparable calculation without this symmetry. Due
to the model spontaneously breaking the $\mathrm{SU}(2)$ symmetry in
the ground state the obtained energies are not comparable and the data
is included here only for completeness.

\paragraph{The error in energy} per spin with and without symmetries
is given in Fig.~\ref{fig:sq:error-d} relative to the reference value
$E_0 \equiv -0.669437$ \cite{sandvik97:_finit_heisen}. The calculation
is initialised in a Néel state ($D=1$) or (for the $\mathrm{SU}(2)$
case) two singlets ($D=2$) and then proceeds in stages. We first
increase the bond dimension to $d=2$ and then time evolve with step
size $\delta\tau = 1, 0.1, 0.01$ until the energy (estimated every
five steps) stops decreasing. Subsequently, we either decrease
$\delta\tau$ or, if we reached the minimal specified $\delta\tau$,
increase $d$ and reset $\delta\tau$ to a larger value
($0.1$). Starting at $d = 3$, we allow a smaller minimal
$\delta\tau = 0.001$. The Trotter error should hence be of order
$O(\delta\tau^2) \approx 10^{-6}$. The energy estimates during the
time evolution are at a fixed $\chi$ and with a re-initialised
environment to trade-off accuracy and computational effort. Each
obtained state is however also saved to disk.

The expectation values shown are then evaluated on the stored states
as described in Sec.~\ref{sec:impl-details}, in particular, error bars
estimate the error of the calculated value compared to the true value
of the given state. Due to our maximal environment bond dimension
being limited to $\chi = 128$, error bars become relatively large in
the logarithmic plot. Data for $\mathrm{SU}(2)$ is only included for
completeness, so let us focus on the three other curves for
calculations with no symmetries, $\mathbb{Z}_2$ symmetry and
$\mathrm{U}(1)$ symmetry:

Contrary to Ref.~\cite{bauer11:_implem_abelian} but in agreement with
Refs.~\cite{liu15:_simpl_heisen, haghshenas18:_j_heisen} we find very
comparable results with and without symmetries at identical bond
dimensions. In particular at larger bond dimensions ($D=5,6$),
enforcing the symmetry appears to aid the calculation as we can guide
the state into the correct sector. This difference is most likely an
effect of the fixed quantum number distribution with an equal number
of states per sector used in Ref.~\cite{bauer11:_implem_abelian} which
reduces the expressive power of the state.

As expected, growing computational effort makes calculations at larger
$D$ only accessible with the full $\mathrm{U}(1)$ symmetry. In
addition to the reduced matrix sizes, the blocking of tensors also
allows for a straightforward shared-memory parallelisation which was
used here to obtain converged $D \geq 7$ results at acceptable
wallclock times of no more than two days. For these larger bond
dimensions, we observe that the obtained error gradually ceases to
shrink. We expect that a variational update of the tensors should
decrease the energy further.

\subsection{The Kagomé lattice}

\begin{figure}[tb]
  \begin{minipage}[t]{0.49\textwidth}
    \includegraphics[width=\textwidth]{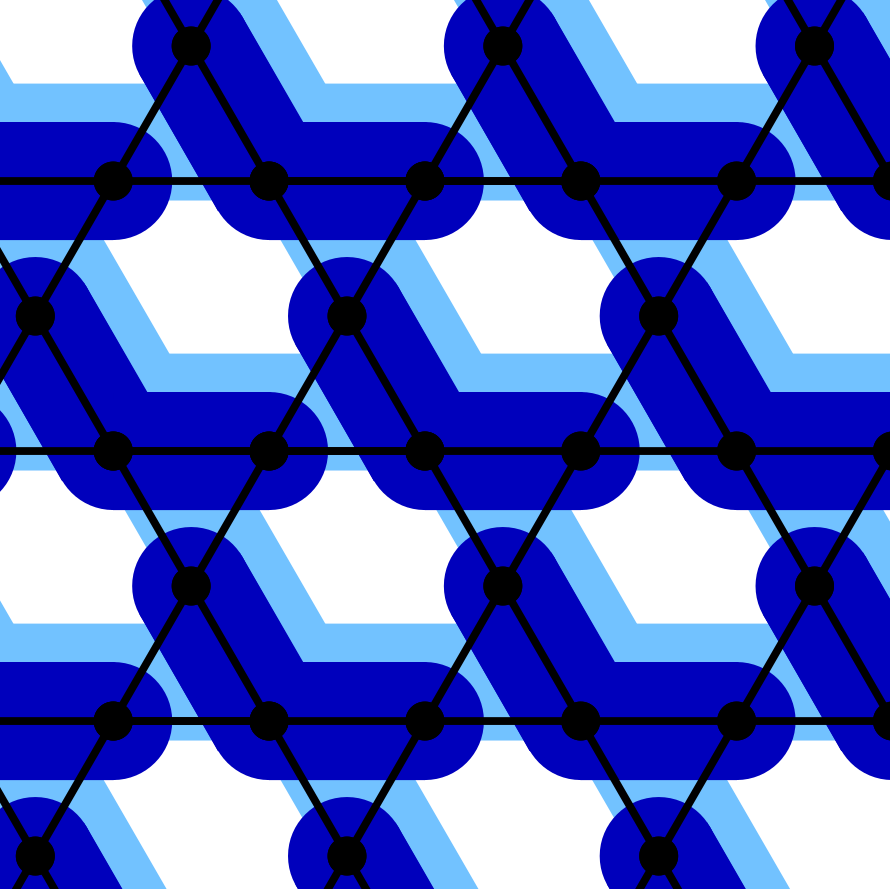}
    \caption{\label{fig:kag:lattice}The mapping of the Kagomé lattice
      to the square lattice used here. Three sites of the Kagomé
      lattice (black circles) are merged into one site of the square
      lattice (dark blue). Each link between two square-lattice sites
      (light blue) corresponds to two links of the Kagomé lattice
      (black lines).}
  \end{minipage}
  \hspace{0.02\textwidth}
  \begin{minipage}[t]{0.49\textwidth}
    \includegraphics[width=\textwidth]{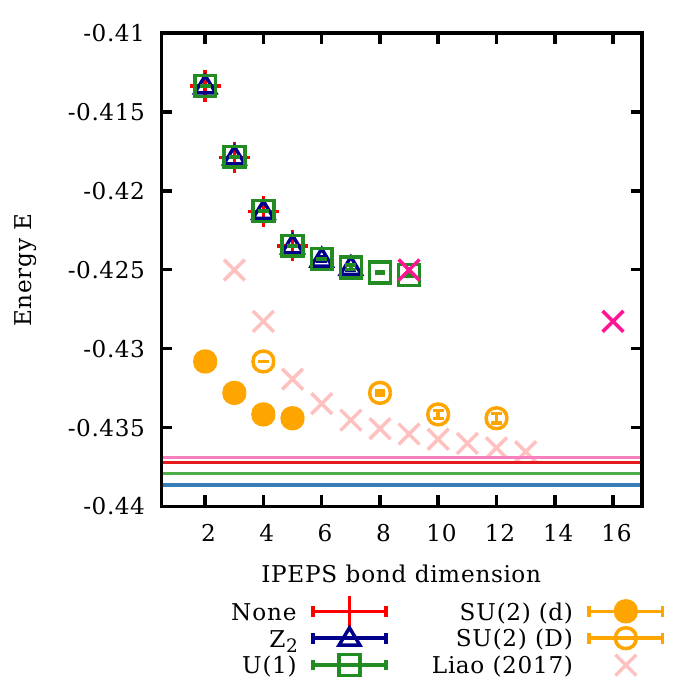}
    \caption{\label{fig:kagome:energy-dx}Ground state energy per spin
      on the Kagomé lattice. The square lattice mapping leads to
      energies which are considerably worse than the current best
      estimates $E_0 \approx 0.437$ (values from
      \cite{mei17:_gapped_heisen},
      \cite{liao17:_gaples_spin_liquid_groun_state_kagom_antif},
      \cite{yan11:_spin_liquid_groun_state_s},
      \cite{depenbrock12:_natur_spin_liquid_groun_state} shown as
      lines in top-down order, 9-site PESS from
      Ref.~\cite{liao17:_gaples_spin_liquid_groun_state_kagom_antif}
      as pink crosses).}
  \end{minipage}
\end{figure}

While the Heisenberg Hamiltonian on the Kagomé lattice is the subject
of ongoing research
(e.g.~cf.~Refs.~\cite{depenbrock12:_natur_spin_liquid_groun_state,
  he17:_signat_dirac_cones_dmrg_study, corboz12:_simpl_su_heisen,
  mei17:_gapped_heisen,
  liao17:_gaples_spin_liquid_groun_state_kagom_antif,
  yan11:_spin_liquid_groun_state_s}), here, we only wish to benchmark
the effect of a $\mathrm{SU}(2)$ spin symmetry which is probably not
spontaneously broken in the ground state. Our implementation currently
only handles square lattices, requiring a mapping of the Kagomé
lattice to the square lattice (cf.~Fig.~\ref{fig:kag:lattice} and
e.g.~Ref.~\cite{corboz12:_simpl_su_heisen}). This mapping is exact,
the resulting $8$-dimensional space of the square lattice sites is not
truncated.\footnote{One could, of course, consider such a truncation
  in future works, e.g. excluding the high-spin states
  $S=\nicefrac{3}{2}$ completely from the calculation.} However, the
mapping is not optimal as two links in the Kagomé lattice are
represented by just a single link in the square lattice, which then
requires an approximately squared bond dimension to represent the same
amount of entanglement. This difficulty should affect the calculations
with and without symmetries mostly in the same way and hence does not
preclude an analysis of the effect of the symmetry implementation
here. A smarter approach is to adapt the structure of the tensor
network to the Kagomé lattice and potentially introducing additional
tensors to aid in convergence \cite{corboz12:_simpl_su_heisen,
  mei17:_gapped_heisen,
  liao17:_gaples_spin_liquid_groun_state_kagom_antif}.

Earlier work\cite{mei17:_gapped_heisen,
  liao17:_gaples_spin_liquid_groun_state_kagom_antif} has found a
ground state energy of approximately $-0.437$ using iPEPS calculations
with the simple update mechanism. DMRG calculations on cylinders
extrapolated to the 2D thermodynamic limit resulted in ground state
energies per spin of
$-0.4386(5)$\cite{depenbrock12:_natur_spin_liquid_groun_state} if
$\mathrm{SU}(2)$ symmetry was conserved and $-0.4379(3)$ without
conservation of $\mathrm{SU}(2)$
symmetry\cite{yan11:_spin_liquid_groun_state_s} and subsequently a
smaller effective bond dimension. 

Fig.~\ref{fig:kagome:energy-dx} shows the results for the energy per
spin with and without symmetries for our implementation. The figure
also includes reference values as follows: first, solid lines show the
extrapolated, final values from Ref. \cite{mei17:_gapped_heisen},
\cite{liao17:_gaples_spin_liquid_groun_state_kagom_antif},
\cite{yan11:_spin_liquid_groun_state_s} and
\cite{depenbrock12:_natur_spin_liquid_groun_state} (in top-down
order); the first two are iPEPS calculations whereas the latter two
are DMRG calculations on cylinders extrapolated to the thermodynamic
limit. Second, light pink crosses reproduce the 9-site PESS data in
Ref.~\cite{liao17:_gaples_spin_liquid_groun_state_kagom_antif}. Third,
the two bright pink crosses on the right-hand side show the same data,
but for an effectively squared bond dimension as
Ref.~\cite{liao17:_gaples_spin_liquid_groun_state_kagom_antif} does
not use the square lattice mapping.

Calculations with no symmetries, $\mathbb{Z}_2$ symmetry and
$\mathrm{U}(1)$ symmetry obtain essentially identical results with the
only difference being a drastic computational speed-up allowing $D=5$
without symmetries but $D=9$ with $\mathrm{U}(1)$ symmetry. This is
expected and fully in-line with calculations on the square lattice
above. Given our mapping of the Kagomé to the square lattice, also the
difference between the results here and earlier works is not too
worrisome. Even a certain degree of continuity between our data and
the data from
Ref.~\cite{liao17:_gaples_spin_liquid_groun_state_kagom_antif} plotted
over the squared bond dimension (bright pink crosses) is visible. As
the square lattice bonds need to transport approximately twice as much
entanglement as the Kagomé lattice bonds in
Ref.~\cite{liao17:_gaples_spin_liquid_groun_state_kagom_antif}, this
is also expected.

The comparison with the $\mathrm{SU}(2)$-invariant calculation (yellow
circles in Fig.~\ref{fig:kagome:energy-dx}) is more interesting:
Comparing both reduced and total bond dimensions, we obtain
substantially lower energies with $\mathrm{SU}(2)$-invariant tensors
than without. While the former is to be expected, the latter is
surprising: Given two tensors of the same total bond dimension $D$,
both are able to represent the same state and hence can be used to
construct states with identical energies.

Instead, the update procedure used here (fast full update) and its
associated convergence problems must be responsible for the drastic
difference in energies: it appears to be much easier to obtain a
low-energy state using $\mathrm{SU}(2)$-invariant tensors than using
unconstrained tensors. Multiple effects likely contribute: first, the
initial random state with $\mathrm{SU}(2)$ invariance is typically a
much better ground-state candidate than a random state which is not
$\mathrm{SU}(2)$ invariant. In the latter case, the imaginary time
evolution first has to rebuild the $\mathrm{SU}(2)$ invariance of the
state from an initially symmetry-broken state. Second, during this
reconstruction, the truncation of the state is not guaranteed to
preserve its (attempt at) $\mathrm{SU}(2)$ invariance. Instead,
degenerate multiplets may be truncated in the middle to obtain the
target bond dimension. Third, while the gauge fixing mechanism used
here greatly improves the accuracy and efficiency of the calculation,
it is still not perfect and large bond dimensions may still lead to
relatively large condition numbers of the norm tensors. By only
considering smaller reduced tensors, this problem is avoided longer in
the $\mathrm{SU}(2)$ invariant calculation.

\subsection{Summary for the computational speed-up}

In this section, we presented results showing that by implementing
either abelian or non-abelian symmetries, larger bond dimensions and
hence better results are obtainable at comparable computational
effort. In the Heisenberg model, $d=10$ is possible when implementing
the $\mathrm{U}(1)$ symmetry, which approximately doubles the bond
dimension compared to our best calculations without any symmetries. As
expected, the $\mathbb{Z}_2$ symmetry provides a smaller benefit than
the $\mathrm{U}(1)$ symmetry but still a noticeable speed-up. In the
Kagomé model, implementing the $\mathrm{SU}(2)$ symmetry leads to much
faster convergence and hence much better results compared to
calculations without symmetries.

We must stress that these are ``merely'' practical advantages: given a
sufficiently fast computer, there is no computational benefit in
implementing a symmetry into a tensor network. However, we will show
in the next section that implementing or not implementing a symmetry
provides an additional control which can be used to obtain
qualitatively different results.

\section{Diagnosing spontaneous symmetry breaking}\label{sec:sym-breaking}

Implementing a symmetry in a system not only allows for a potential
computational speed-up, but also provides additional measures to
control the calculation.

First, forcing a state to transform uniquely under a symmetry is
equivalent to doing the calculation in a canonical instead of
grand-canonical ensemble. As a result, it becomes possible to fix
total particle number (or average particle number per unit cell)
exactly to a desired value. For example in the Kagomé calculation, the
states with $\mathrm{SU}(2)$ symmetry are exact eigenstates of the
total spin operator with eigenvalue $S \equiv 0$ and the states with
$\mathrm{U}(1)$ symmetry are exact eigenstates of the total $\hat S^z$
operator with eigenvalue $S^z \equiv 0$. In contrast, the states
without any symmetry only have $S^z \approx 0$. It then becomes
possible to e.g.~exactly fix the total particle number per unit cell in the
Hubbard model instead of relying on a chemical potential to obtain the
desired filling fraction. On the flip side, filling fraction and unit
cell size have to be commensurate to provide an integer average
particle number per unit cell. This integer average particle number is
removed by the ``remover'' tensor described in
Sec.~\ref{sec:impl-details:adaptations} and Fig.~\ref{fig:remover}.

Second, symmetry-preserving tensors can only represent eigenstates of
the symmetry. If the symmetry operator and the Hamiltonian commute and
are non-degenerate, each eigenstate of the Hamiltonian is also an
eigenstate of the symmetry operator and this is no restriction. If,
however, the Hamiltonian has a degenerate spectrum, eigenstates of the
Hamiltonian are not necessarily eigenstates of the symmetry
operator. Then optimising eigenstates of the symmetry operator for the
lowest energy does not yield a natural eigenstate of the Hamiltonian
but instead some superposition of degenerate eigenstates. Such
superpositions have typically larger entanglement, or, if we fix the
maximal entanglement, a higher energy than if we allowed breaking of
the symmetry in the ansatz.

The increased entanglement of symmetric superpositions can be used to
diagnose spontaneous symmetry breaking: If a symmetry is not broken,
then it should be possible to implement this symmetry in the tensor
network. The resulting state should be an equally good candidate to
represent the ground state as a state constructed from unrestricted
tensors. As such, at a fixed bond dimension, we expect to find the
same variational energy regardless of whether we implement an abelian
symmetry or not provided that there are no problems with convergence
and the symmetry sectors are chosen in an optimal fashion. Non-abelian
symmetries are more tricky, as they increase the effective bond
dimension of the state and hence lead to more difficult
comparisons. If, on the other hand, a symmetry is spontaneously
broken, we expect a better variational energy if we also allow the
tensors in our tensor network state to break this symmetry. For lack
of spontaneous symmetry breaking in finite and also one-dimensional
infinite systems, this is not a problem often encountered. In iPEPS
calculations, however, it does become relevant.

\begin{figure}[tb]
  \begin{minipage}[t]{0.49\textwidth}
    \includegraphics[width=\textwidth]{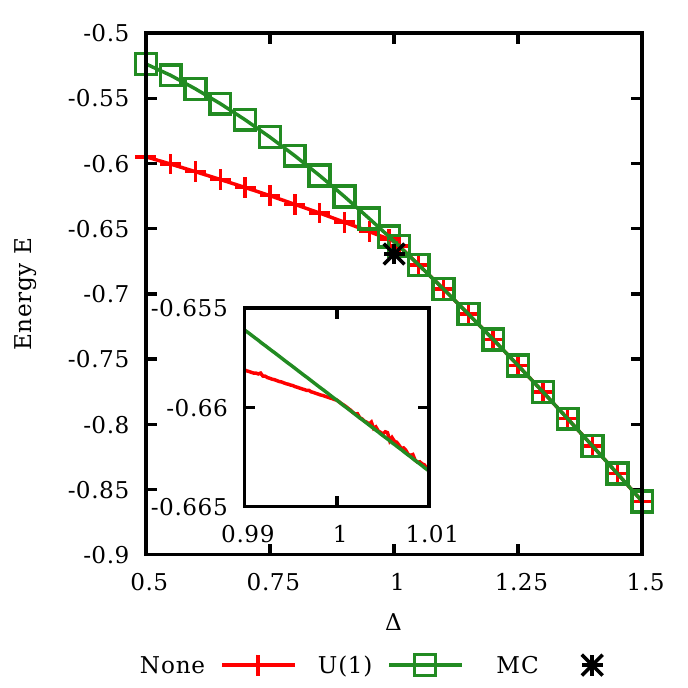}
    \caption{\label{fig:diagnose:heisenberg-d2}Obtainable ground-state
      energies in the anisotropic Heisenberg model at $D = 2$ with and
      without $\mathrm{U}(1)$ symmetry. Inset: Zoom into the central
      region, $\Delta$ values are placed $0.0025$ apart. Lines are
      guides to the eye. The results start to differ at $\Delta <
      1$.}
  \end{minipage}
  \hspace{0.02\textwidth}
  \begin{minipage}[t]{0.49\textwidth}
    \includegraphics[width=\textwidth]{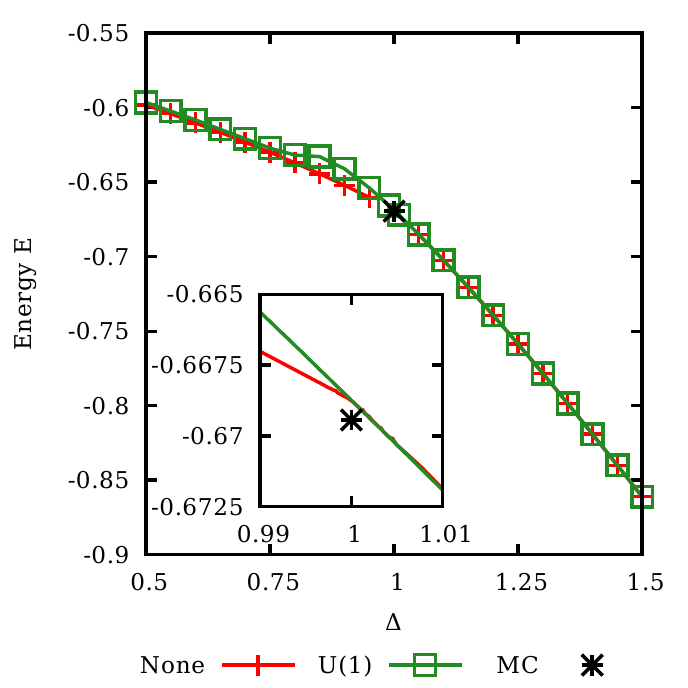}
    \caption{\label{fig:diagnose:heisenberg-d4}As left, for $D=4$. At
      small enough $\Delta$, the $\mathrm{U}(1)$ symmetric state
      becomes competitive again but still has slightly higher energy
      than the unconstrained state. Monte Carlo reference value at
      $\Delta = 1$ taken from
      Ref.~\cite{sandvik97:_finit_heisen}. Error bars are smaller than
      symbol size in all cases.}
  \end{minipage}
\end{figure}

\paragraph{The Heisenberg model on the square lattice} as discussed
before can serve as an example of this effect. The Hamiltonian is:
\begin{equation}
  \hat H = \sum_{\langle i, j \rangle} \hat s^x_i s^x_j + \hat s^y_i \hat s^y_j + |\Delta| \hat s^z_i \hat s^z_j \quad.
\end{equation}
with now $\Delta$ not fixed at $1$. At $\Delta < 1$, the ground state
spontaneously breaks the $\mathrm{U}(1)$-$S^z$ symmetry and only keeps
a canted order. If, for a fixed and finite bond dimension $D$, we then
plot the optimal energy obtained over the value of $\Delta$ for
different calculations with and without that symmetry, we can
straightforwardly diagnose the transition point,
cf.~Figs.~\ref{fig:diagnose:heisenberg-d2} and
\ref{fig:diagnose:heisenberg-d4}. Here, we only need to measure the
energy of the system -- a local observable which is straightforward to
evaluate with reliable error bars. More importantly, states are
optimised for their energy. The energy is hence typically the
observable which behaves in the most stable manner and most closely
represents the value obtained in an actual physical state. By
comparing the behaviour of the different calculations over the
transition point at $\Delta = 1$, we can easily conclude that the
$\mathrm{U}(1)$-$S^z$ symmetry is spontaneously broken at $\Delta < 1$
due to the considerably lower energy obtained from an unconstrained
calculation.

Furthermore, while a clear kink at $\Delta = 1$ is visible in the
energy in calculations without symmetry, this kink is absent and the
energy a nearly linear function of $\Delta$ around $\Delta = 1$ in the
calculation with $\mathrm{U}(1)$ symmetry. The strong bias towards a
symmetry-unbroken state completely hides the symmetry-broken phase
from the $\mathrm{U}(1)$-symmetric calculation. Hence, by implementing
a symmetry in a system, we can not only select a sector in which we
want to do our calculation (e.g. $S^z = 0$ or $N/L = 0.5$) but also --
to some extent -- select the symmetries present in the calculated
states. Of course, at abitrarily large $D$, it becomes possible for
the calculation to hide the symmetry-breaking by a suitable
superposition as discussed before. This effect can be seen at
$\Delta < 0.8$ and $D \geq 4$, where the $\mathrm{U}(1)$-symmetric
calculation is again somewhat comparable to the unsymmetric
calculation.

\paragraph{If we implement the $\mathbb{Z}_2$ symmetry,} we disallow
unit cell states with $ S^z = \pm 1$, on the $2 \times 2$ unit cell in
a vacuum environment this hence limits us to the two ferromagnetic
states (which are very highly energetic and don't play a role) and the
six $ S^z = 0$ states. With a suitable environment, one could
alleviate this problem by combining a $ S^z = 1$ state from the
environment with a local $S^z = -1$. However, introducing such a state
would break the $\mathrm{U}(1)$ symmetry (as it should), which is
\emph{preserved by the Hamiltonian we use for the
  evolution}.

Put another way, starting from an antiferromagnetic state
$|\psi\rangle$ with $S^z = 0$, the global ground state is not
contained in the Krylov subspace of $\hat H$ and $|\psi\rangle$. As
such, it is not possible to obtain it by imaginary time evolution
alone. Worse, if we initialise the evolution with a symmetry-broken
state (of comparatively high energy), the symmetry-broken contribution
to this state is very quickly removed by the automatic truncation and
hence not available when it would be useful to lower the energy. This
problem is not limited to calculations with $\mathbb{Z}_2$
symmetry. Also for calculations without any symmetries, it was
necessary to select multiple random initial states and run the
calculation at $D=2$ multiple times to obtain the correct ground-state
energy. Initialising the calculation without symmetries with a Néel
state resulted in energies typically very close to those obtained from
the calculation with full $\mathrm{U}(1)$ symmetry. Only very rarely
have we found the error from the truncation to lead to spontaneous
symmetry breaking by itself.

Avenues to avoid this problem could be random noise terms added to the
state or special ``tunneling'' Hamiltonians
(e.g.~$\hat s^+ \hat s^+ + \hat s^- \hat s^-$) which are added to the
Hamiltonian with a small prefactor and allow for tunneling between
different symmetry sectors. The latter approach is already fairly
common when solving complicated systems using DMRG. This, however, is
outside the scope of the present paper.

\section{Conclusion}\label{sec:conclusions}

We have shown a considerable computational advantage to be derived
from the implementation of symmetries in iPEPS tensor networks. For
example implementing a $\mathrm{U}(1)$-$S^z$ symmetry in the
Heisenberg square lattice model allows for a maximal bond dimension
$D=10$ at effort comparable to a calculation without any symmetries at
$D=6$. Furthermore, we show that the explicit Clebsch-Gordan
Coefficient approach is sufficient to represent
$\mathrm{SU}(2)$-symmetric tensors and provides for a large advantage
in the Kagomé Heisenberg model. Finally, we provide an example of how
implementing a symmetry allows for further control of the calculation
by firstly either enabling or disabling the symmetry itself and
secondly selecting the desired quantum number sector if the symmetry
is enabled.

\section*{Acknowledgements}
We are thankful for very helpful discussions with I. P. McCulloch,
M. C. Ba\~{n}uls, J. I. Cirac, L. Vanderstraten, F. Verstraete and
N. Schuch. We also much appreciated comments by E.~M.~Stoudenmire on
the first version of this paper.

\paragraph{Funding information}
Funding through ERC Grant QUENOCOBA,
ERC-2016-ADG (Grant no. 742102) is acknowledged.


\end{document}